# Observation of robust macroscale structural superlubricity

Minhao Han[*,1], Deli Peng[*#,1,2], Dinglin Yang[3,4], Jin Wang[5], Yi Zheng[2], Guofeng Hu[2], Meng Qi[2], Yifan Shao[1], Jiaying Li[6], Feng Ding[7], Zhiping Xu[3,4], Michael Urbakh[8], and Quanshui Zheng[#,1,2,3,4]

[1]*Center of Double Helix, Institute of Materials Research, Tsinghua Shenzhen International Graduate School, Tsinghua University, Shenzhen, China.*
[2]*Institute of Superlubricity Technology, Research Institute of Tsinghua University in Shenzhen, Shenzhen, China.*
[3]*Center for Nano and Micro Mechanics, Tsinghua University, Beijing, China.*
[4]*Department of Engineering Mechanics, Tsinghua University, Beijing, China.*
[5]*International School for Advanced Studies (SISSA), Trieste, Italy.*
[6]*College of New Materials and New Energies, Shenzhen Technology University, Shenzhen, China.*
[7]*Suzhou Laboratory, Suzhou, China.*
[8]*School of Chemistry and The Sackler Center for Computational Molecular and Materials Science, Tel Aviv University, Tel Aviv, Israel.*

[*]Equal contribution: Minhao Han; Deli Peng.
[#]Corresponding author. Email: pengdeli@sz.tsinghua.edu.cn; zhengqs@tsinghua.edu.cn.

**ABSTRACT**. Structural superlubricity (SSL) promises nearly frictionless and wearless sliding, but has until now been considered a special and extreme interfacial phenomenon limited to micro- and nanoscale contacts. Here, we demonstrate robust macroscale SSL within a single sub-millimeter graphite contact. Previously reported near-zero friction coefficients, where friction is nearly independent of normal load, have only been observed at microscale contacts under low loads. Our system expands both contact size and load into the macroscopic regime, exhibiting friction coefficients that fluctuate around zero and reach values as low as $\sim 10^{-6}$ across a broad load range from 1 mN to 0.5 N. Negative friction coefficients are also observed. Similar behavior is observed at graphite/$MoS_2$ interfaces, indicating that macroscale SSL is a generalizable phenomenon across flat layered materials. These findings overturn long-standing scaling limitations and establish macroscale SSL as a paradigm-shifting platform for next-generation mechanical and electromechanical systems.

## I. INTRODUCTION.

Relative sliding between two contacting solid surfaces is one of the simplest forms of constrained motion in the material world, occurring across a wide range of length and time scales. Friction and wear play central roles in this process, leading to significant energy losses and material consumption [1,2]. Structural superlubricity (SSL), a state of nearly frictionless and wearless sliding between solid crystalline surfaces, presents a revolutionary solution for overcoming friction and wear challenges [3-5]. However, theoretical, numerical, and experimental studies of SSL remain limited to the nano- and microscale [6-12], hindered by factors such as intrinsic elasticity [10-14], lattice defects and unwanted chemical bonding [3,7,15,16], grain boundaries [17-19], and surface contamination [20,21]. Although SSL-induced self-retraction behavior has been observed in centimeter-scale carbon nanotubes [22], the contact area remains nanoscale.

A promising strategy for constructing macroscale superlubric (SL) interfaces relies on the multi-contact paradigm, which involves breaking down a macroscale interface into a large set of nanoscale incommensurate SSL contacts that are either independent or weakly coupled [4,23,24]. Examples include graphene-wrapped nanoscale spheres [25,26], ordered 2D flake junctions [27,28], or the recently synthesized graphullerene (GF) carbon allotrope, also known as quasi-hexagonal phase fullerene [29,30]. While these approaches enable very low friction and are relatively easy to integrate into engineering systems, they still suffer from non-negligible energy dissipation due to edge and corner pinning, dynamic contact formation and rupture [3,4,31], stress-induced defects, and strong sensitivity to humidity [25,28]. Consequently, the lowest friction coefficient reported for macroscale SL ($\sim 10^{-3}$) remains orders of magnitude higher than that of microscale SSL [32,33]. Extending single-contact SSL from the microscale to the macroscale offers the potential for much lower dissipation and wearless operation. While microscale SSL has been demonstrated using naturally incommensurate twist interfaces in highly oriented pyrolytic graphite (HOPG) [7], the limited grain size and presence of internal defects in HOPG prevent this strategy from scaling up [34].

In this work, we have developed a novel method for fabricating submillimeter-scale, nearly defect-free, atomically smooth single-crystal graphite sliders, leveraging recently reported epitaxial growth of single-crystal graphite [35]. We further demonstrate the persistence of robust single-contact structural superlubricity at the macroscale.

## II. RESULTS

Centimeter-scale single-crystal graphite films were grown by continuous epitaxy through isothermal carbon diffusion through nickel (FIG. S1), following previously reported methods [35]. Patterned square graphite mesas with gold caps (0.02–0.2 mm) were fabricated by photolithography and lift-off (FIG. S1). To achieve clean and incommensurate single-crystal graphite interfaces, we developed a cleavage-transfer-stacking technique. As shown in FIG. 1(a), the sample was mounted on a piezoelectric stage. Polydimethylsiloxane (PDMS) protrusions were prepared using a molding method (FIG. S2). Because the adhesion between PDMS and the mesa cap is normally insufficient for cleavage, the sample was rapidly pulled using the piezo stage to exploit the viscosity-dependent adhesion of PDMS, enabling instant cleavage between graphite layers. This process produced freshly cleaved, patterned graphite sliders that adhered to PDMS (FIG. 1(a)). Optical, Scanning Electron Microscope (SEM), and atomic force microscope (AFM) characterizations showed that this method yields ultra-clean, uniform, defect-free, atomically smooth surfaces on the epitaxial single-crystal graphite slider, with no steps, wrinkles, or other defects that may eliminate SSL (FIGs. 1(b)–(c)). Electron back scatter diffraction (EBSD) analysis confirmed that the entire surface of the slider remained single-crystalline (FIG. 1(d)). These results demonstrate the successful preparation of nearly perfect single-crystal graphite sliders, visible to the naked eye, as highlighted by the size comparison with a human hair (FIG. 1(b)).

Clean single-crystal graphite substrates with large flat regions were also prepared on silicon wafers by mechanical exfoliation (FIG. S3). AFM and EBSD confirmed these surfaces to be atomically smooth and single-crystalline. Finally, as shown in FIG. 1(a), the slider on PDMS was rotated and released onto the graphite substrate by precise piezo control, forming an incommensurate stacking state. The largest contact area achieved was 0.2 mm × 0.2 mm, over two orders of magnitude larger than previously reported nanoscale flakes and microscale HOPG mesas [5,7].

To probe the contact interface, we used focused ion beam (FIB) milling to section the sliding region. As shown in the cross-sectional SEM image (FIG. 1(e), left), the graphite slider and substrate are clearly distinguishable at the side edge. In the inner contact region, no contrast variations or visible gaps are observed between them through SEM (FIG. 1(e), top right), indicating intimate contact. High-resolution scanning transmission electron microscopy (STEM) imaging (FIG. 1(e), bottom right) reveals diffraction contrast arising from the different crystal orientations of the slider and substrate. In the incommensurate stacking state, this contrast allows clear identification of the sliding interface, marked by a white dotted line. No signs of contamination or diffusion from the surrounding regions are observed at the interface [36]. Applying an inverse Fourier transform to the STEM image produces a selected area electron diffraction (SAED) pattern showing two distinct single-crystal lattice sets (FIG. 1(f)). The measured ~11° twist angle between them confirms the incommensurability of the contact. These results demonstrate that the macroscopic sliding interface we have constructed is nearly in full atomic contact.

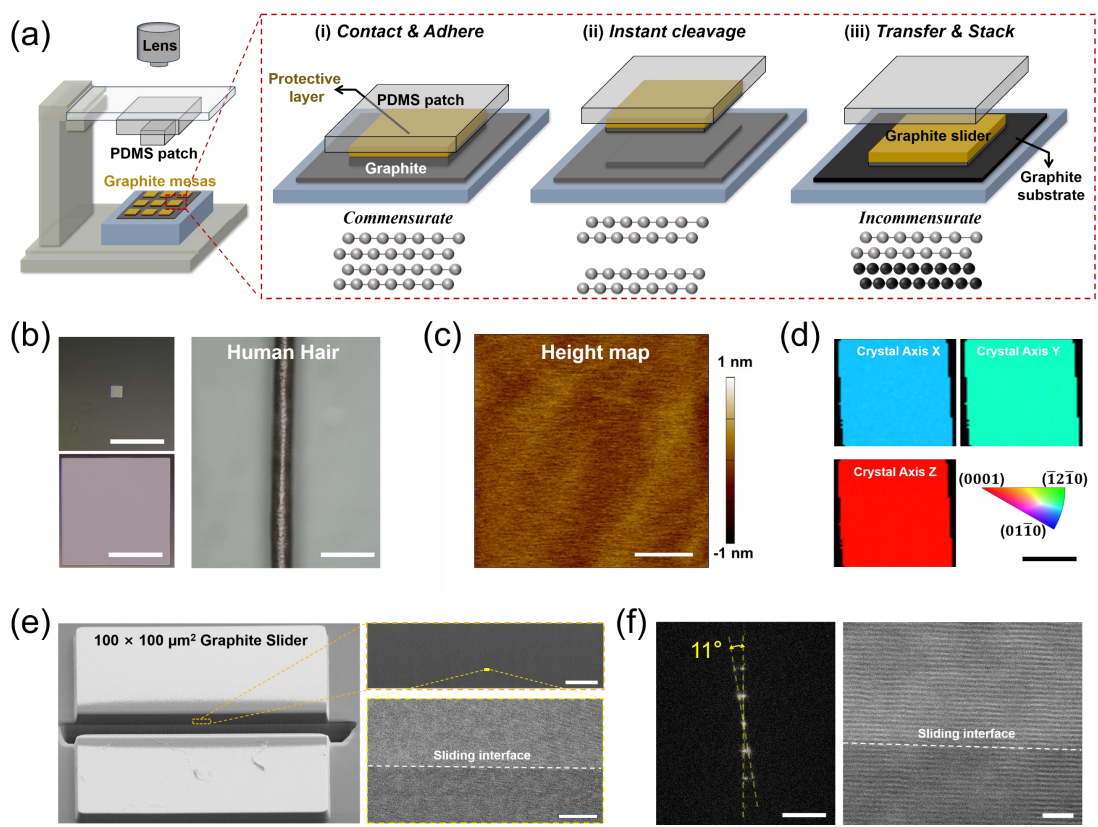

FIG.1 Construction and characterization of the macroscale, full atomic contact van der Waals sliding interface. (a) Schematic of cleavage and transfer of a large graphite slider. (b) Optical images of PDMS-cleaved graphite mesas with sizes of 0.02 mm (top left) and 0.2 mm (bottom left); right image shows a human hair for scale. Scale bar, 0.1 mm. (c) AFM image of the bottom surface of a 0.1 mm mesa. Scale bar, 5 μm. (d) EBSD image of the bottom

surface of a 0.1 mm mesa, confirming its single-crystalline nature. Scale bar, 50 μm. (e) FIB-sectioned SEM image of 0.1×0.1 mm$^2$ slider-substrate interface. Scale bar, 1 μm. Bottom right: STEM image of selected interface. Scale bar, 20 nm. (f) STEM images of the same interface. Left: SAED pattern gives ~11° twist angle between slider and substrate. Scale bar, 5 1/nm; Right: magnified STEM image with the interface highlighted by a white dashed line based on diffraction contrast, indicating full atomic contact. Scale bar, 2 nm.

To investigate the frictional properties of these graphite interfaces, we developed a visualized friction testing system, as shown in the inset of FIG. 2(a). This system uses a 5-mm-diameter glass hemisphere attached to a biaxial force sensor to apply a load to the graphite slider. A displacement stage beneath the substrate provides reciprocal motion between the slider and the graphite substrate while recording friction forces in real time. In all experiments, the sliding velocity was fixed at 10 μm/s. An optical microscope enables real-time observation of the slider's movement.

FIG. 2(a) shows the load test results for a 0.1 mm square slider. Across a load range from 1 mN to 0.5 N, the friction force remained extremely low, measuring approximately 5.12 μN under a 0.5 N load, corresponding to an average shear stress ($\sigma_s$) of only 0.51 kPa. Based on the Newton ring pattern in FIG. S4, the contact area between the ball and the upper surface of the slider was estimated to be approximately 204.5 μm$^2$. At a normal load of 0.5 N, the average pressure at the center of the slider's upper surface was calculated to be ~2.45 GPa. This normal pressure is sufficient to handle most application scenarios.

Throughout the tested load range, the friction force was effectively independent of the applied normal load, with a fitted differential coefficient of friction (COF) as low as ~1.73 × 10$^{-6}$. For the largest 0.2 mm square sliders, the experiments also demonstrated load-independent friction (FIG. S5), with a fitted COF of ~8.85 × 10$^{-6}$. We also observed negative friction coefficients in many samples. FIGs. 2(b)–(c) show two representative cases, with the COF values of -4.54 × 10$^{-6}$ and -5.01 × 10$^{-5}$, respectively. The influence of slider thickness on the load dependence of friction was also examined, and the results indicate that its effect is very weak within the thickness range of the sliders used in our study (Fig. S9).

The negative friction coefficients may be attributed to load-induced suppression of out-of-plane motion of edge atoms or moiré ridges, where increasing normal load leads to reduced energy dissipation [37], or to load-induced edge detachment, which lowers the friction force at the edges [38]. Regardless, this is the first report of anomalous negative friction coefficients observed in macroscale tribological experiments.

The fact that the COF fluctuates between slightly positive and negative values further suggests that the true COF of our system is extremely close to zero, with competing interfacial mechanisms contributing to minor deviations on either side. We attribute this near-zero COF to the nearly full atomic-contact nature of the submillimeter-scale van der Waals interface, in which even under significantly increased normal load the contact area remains unchanged and no obvious alteration in interfacial interactions occurs.

In addition, when rotating these macroscopic sliders, we observed self-locking behavior (Fig. S6), which has previously been demonstrated at the nano- and microscale [5,7]. In this case, the interface friction transitions from an ultralow state to a very high state such that the slider cannot even be moved by our glass ball.

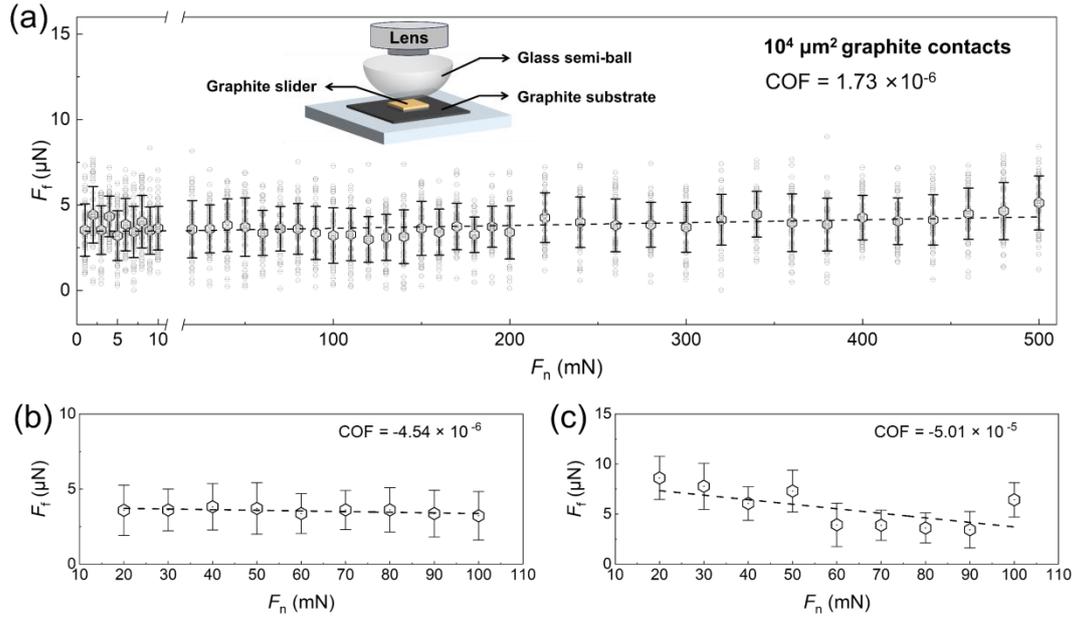

**FIG.2** Load dependence of friction at macroscopic SSL interfaces. (a) Friction force as a function of applied normal load (1 mN to 0.5 N) for $0.1 \times 0.1$ mm$^2$ incommensurate graphite contacts. Black hollow hexagons and vertical bars represent the mean and standard deviation of the friction force, respectively, measured over 50 sliding cycles under each normal load. The black dashed line shows the fitted curve for the mean friction force as a function of load. (b, c) Friction force versus applied normal load for two additional samples with $0.1 \times 0.1$ mm$^2$ incommensurate graphite contacts, both exhibiting negative friction coefficients.

By measuring the friction force across the incommensurate graphite interfaces with contact areas ranging from 400 μm$^2$ to 40000 μm$^2$, we obtained the scaling relationship with contact size, as shown in FIGs. 3(a)–(b). The friction force exhibits a sublinear dependence on contact area, following $F_f \propto A^{0.39}$ (FIG. 3(a)), which is characteristic of SSL and reflects the effective cancellation of lateral forces across the internal surface area. Our findings are consistent with previous results obtained at the nanoscale for gold clusters sliding on a graphite superlubric interface [12,39], and at a micrometer-scales for graphite/graphite and MoS$_2$/graphite superlubric interface [32,40]. The measured scaling exponents vary across different experimental systems [13], likely due to the influence of such factors as material structure and properties, contact shape, twist angles, sliding directions, presence of lattice defects, as well as environmental and testing conditions. Lower values of scaling exponent indicate better cancellation of resisting lateral forces acting on the inner slider surfaces and its edges. Consistently with the friction force scaling, the interfacial shear stress decreases with increasing area (FIG. 3(b)). These findings clearly demonstrate that for graphite/graphite interfaces, superlubric sliding, previously observed at nano- and micro-scales can be successfully extended to the submillimeter macroscale.

The presented results naturally raise an important question: Can SSL be further scaled to even larger dimensions, and what is the ultimate length scale that defines the limits of structural superlubric contact? Theoretical and numerical studies suggested that factors such as the elasticity of inner contact surfaces and slider edges may impose constrains on the upscaling of SSL [3,13]. Regarding static friction [10], a significant increase in friction and elimination of sublinear scaling with the interface area was predicted for the contact sizes exceeding the characteristic length scale for elastic dislocations. At this point, interfacial deformations localize the misregistry into dislocation cores, and sliding proceeds through dislocation nucleation and motion, leading the frictional shear stress to reach a finite plateau value independent of size. The critical size in this context reads as:

$$D = \frac{G}{\tau_{max}} a$$

where $G$ is the in-plane shear modulus, $\tau_{max}$ is the maximal lateral shear stress experienced by an atom on the surface potential, and $a$ is the lattice constant. Specifically, for graphite contacts, where $G \approx 400$ GPa, $\tau \approx 60$ MPa [41], and $a = 2.46$ Å, the critical size, $D$, is estimated as 1~2 μm — far smaller than the dimensions achieved in our experiments. This discrepancy can be attributed to the limitations of the simulation model [13,42], particularly when applied to rigid materials such as 2D materials.

Elasticity effects are obviously not limited to the inner contact surface and can appear also near edges and corners of sliders, where lower atomic coordination implies higher flexibility. This, in turn, may be manifested in larger out-of-plane motion and enhanced tendency to form commensurate regions, thus enhancing frictional energy dissipation. Atomistic simulations [43,44] showed that kinetic friction in layered material interfaces is dominated by dissipation at the corners and edges up to contact dimensions of $10^{-2}-10^{2}$ μm$^2$, depending on the slider shape, misfit angle, normal load and sliding velocity. As the contact size increase above these threshold dimensions, a transition from sublinear to linear scaling of the friction force with the contact area should occurs, with friction originating from the inner interface becoming the dominant contribution.

Notably, for both static and kinetic friction, the predicted threshold sizes at which the frictional stress no longer decreases with increasing contact area, are significantly smaller than those achieved in our experiments (0.2 mm). This discrepancy highlights the need for new theoretical and numerical frameworks, especially for multilayer graphene systems utilized in our experiments. Within the experimentally accessible size range, our results alleviate concerns that intrinsic elasticity inherently limits the scalability of SSL to macroscopic dimensions.

We further measured the dependence of friction force on normal load for different contact sizes. For all sizes, the COF values remained ultra-low and did not show a scaling relationship with contact area (FIG. 3(c)).

To investigate the material universality of macroscopic SSL, we transferred large-area single-crystal graphite sliders onto single-crystal MoS$_2$ substrates (up to 0.15 mm, shown in FIG. S5). The friction dependence on the normal load measured for 0.1 mm × 0.1 mm graphite/MoS$_2$ contacts is shown in FIG. 3(d), respectively. Once again, the friction force remained extremely low (~1.46 μN), corresponding to an average shear stress of ~0.15 kPa, and was nearly independent of normal load, with COF of ~1.62 × 10$^{-6}$. This confirms that macroscopic SSL can be robustly achieved not only in incommensurate homogeneous graphite contacts but also in heterojunction graphite/MoS$_2$ interfaces.

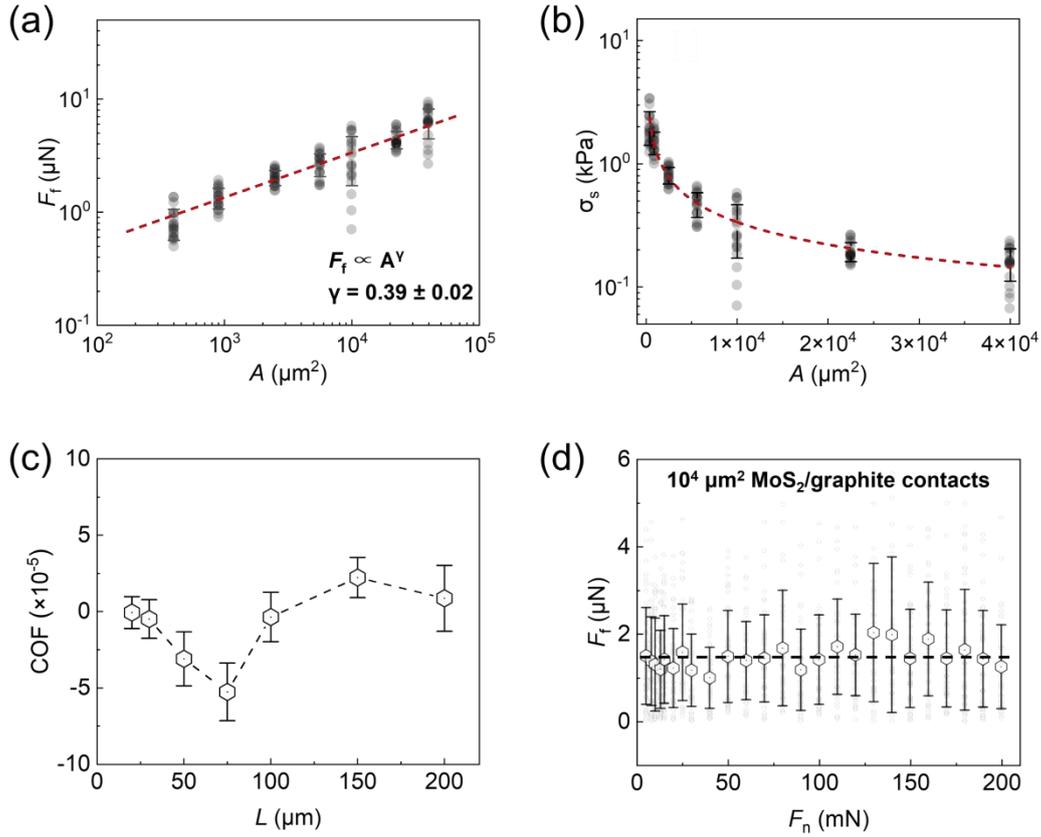

FIG.3 Tribology of macroscopic SSL interface. (a) and (b) Dependence of friction force ($F_f$) and shear stress ($\sigma_s$) on contact area, respectively. The red dashed lines represent the fitting results. (c) COF values as a function of slider length ($L$) of graphite samples. (d) Load dependence of friction force for the $0.1 \times 0.1$ mm$^2$ graphite/MoS$_2$ contact.

From the perspective of engineering, a sliding state with a coefficient of friction (COF) below 0.01 is commonly defined as superlubricity (SL). SL is further categorized as either solid or liquid, depending on whether solid or liquid phase dominates the interfacial mechanics. Reported demonstrations of macroscale SL have involved apparent contact areas ranging from approximately $10^3$ μm$^2$ to $10^6$ μm$^2$. To date [45,46], the lowest COF achieved in solid-SL systems is around $1\times10^{-3}$, while liquid-SL systems have reached values as low as $\sim 2\times10^{-5}$. In the case of SSL, even lower COF down to $\sim 2.3\times10^{-6}$ have been measured [32], but only at nano- to few-micrometer scales. As shown in FIG. 4, our work demonstrates a similarly ultra-low COF at the macroscopic scale, establishing that such exceptional tribological performance can indeed be realized beyond the micro- and nanoscale and extended to engineering-relevant dimensions through deliberate material design and interface engineering. These results position macroscale SSL as a transformative technology that outperforms all known solid and liquid lubrication strategies. Additionally, as shown in FIG. S8, we compared the secant-derived COF values with those reported in the literature and confirmed that the performance of our macroscale SSL system is also superior.

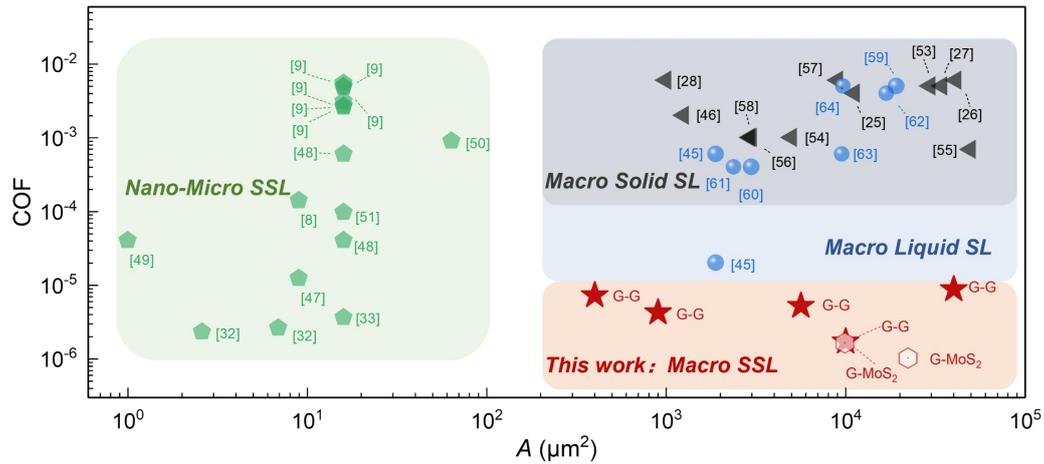

FIG.4 Comparison of friction coefficients of nanoscale/microscale SSL [8,9,32,33,47-51], macroscale solid SL [25-28,46,52-58], macroscale liquid SL [45,59-64] and this work (macroscale SSL). When the contact area ($A$) is not explicitly specified in the article, it is estimated from the scale of the sliding track or by using the Hertz contact model. The friction coefficients of the corresponding friction pairs are all taken as the lowest values reported in each paper. G represents graphite. The data referenced in the FIG. 4 are all in one-to-one correspondence with FIGs. S7-8 in the Supplementary Materials.

## III. DISCUSSION

Since SSL was first experimentally confirmed at nano- to microscale dimensions two decades ago, significant efforts have been devoted to exploring its scalability to macroscopic dimensions. However, this endeavor faces numerous challenges due to the emergence of additional frictional energy dissipation pathways at larger contact scales. These include surface and edge elasticity, grain boundaries, structural defects, interlayer binding and intercalation of contaminants. In this work, we show that our sample preparation and experimental design effectively overcome these limitations and provide direct evidence that robust SSL can persist at the macroscale.

The atomic-level full-contact nature of the weak vdW interface results in an extraordinarily low friction coefficient, where the friction force remains nearly independent of normal load across a wide loading range, markedly different from the behavior of multi-contact configurations. This feature also minimizes the penetration of external contaminants into the sliding interface, further enhancing the stability and durability of the macroscale SSL contact.

The system and methods developed in this study establish a robust foundation for future SSL research at even larger scales, higher loads, and higher sliding speeds—conditions that have been challenging to achieve at the nano- and microscale due to system or instrument limitations. These advancements will provide valuable insights into SSL mechanisms, contribute to the refinement of theoretical models, and pave the way for practical applications in real-world settings.

**Acknowledgments**

We are grateful to Yi Yang, Zirui Yuan, and Miaoxuan Xue for their assistance with the preparation and transfer of epitaxial graphite samples. This work was supported by the National Key R&D Program of China (No. 2023YFB4603604), the National Natural Science Foundation of

China (Grants No. 12574037), the National Key R&D Program of China (No. 2023YFB4603604), Shenzhen Science and Technology Program (No. KQTD20240729102211015, No. JCYJ20210324100600001), Shenzhen Key Laboratory of Superlubricity Technology (No. ZDSYS20230626091701002), the National Natural Science Foundation of Guangdong Province (No. 2025A1515012226).